\documentclass[aps,prl,twocolumn,superscriptaddress,showpacs]{revtex4-1}
\usepackage{graphicx}
\usepackage{dcolumn}
\usepackage{bm}
\usepackage{amsmath,amssymb}
\usepackage[
  colorlinks=true,
  allcolors=blue,
  breaklinks=true
]{hyperref}
\makeatletter
\def\captionof#1#2{{\def\@captype{#1}#2}}
\makeatother

\usepackage[english]{babel}
\usepackage{times}
\usepackage{amsfonts}
\usepackage{psfrag}
\usepackage{verbatim}
\usepackage{color}

\begin{document}


\title{Field-tuned order by disorder in Ising frustrated magnets with antiferromagnetic interactions}

\author{P. C. Guruciaga}
\affiliation{Instituto de Investigaciones F\'\i sicas de Mar del Plata (IFIMAR), 
UNMdP-CONICET, 7600 Mar del Plata, Argentina}

\author{M. Tarzia} 
\affiliation{Universit\'e Pierre et Marie Curie
  - Paris 6, Laboratoire de Physique Th\'eorique de la Mati\`ere Condens\'ee,
  4, Place Jussieu, Tour 13, 5\`eme \'etage, 75252 Paris Cedex 05,  France}

\author{M. V. Ferreyra}
\affiliation{Instituto de F\'{\i}sica de L\'{\i}quidos y Sistemas
Biol\'ogicos (IFLYSIB), UNLP-CONICET, 1900 La Plata, Argentina}

\author{L. F. Cugliandolo} 
\affiliation{Universit\'e Pierre et Marie Curie
  - Paris 6, Laboratoire de Physique Th\'eorique et Hautes Energies,
  4, Place Jussieu, Tour 13, 5\`eme \'etage, 75252 Paris Cedex 05,  France}

\author{S. A. Grigera}
\affiliation{Instituto de F\'{\i}sica de L\'{\i}quidos y Sistemas
Biol\'ogicos (IFLYSIB), UNLP-CONICET, 1900 La Plata, Argentina}
\affiliation{School of Physics and Astronomy, University of St Andrews, 
St Andrews KY16\ 9SS, United Kingdom}

\author{R. A. Borzi}
\email{r.chufo@gmail.com}
\affiliation{Instituto de F\'{\i}sica de L\'{\i}quidos y Sistemas
Biol\'ogicos (IFLYSIB), UNLP-CONICET, 1900 La Plata, Argentina}



\begin{abstract}
We demonstrate the appearance of thermal order by disorder in 
Ising pyrochlores with staggered antiferromagnetic order frustrated by an
applied magnetic field. We use a mean-field cluster variational
method, a low-temperature expansion and Monte Carlo simulations to characterise the 
order-by-disorder transition. By direct evaluation of the
density of states we quantitatively show how a symmetry-broken state is selected by thermal excitations. 
We discuss the relevance of our results to experiments in $2d$ and $3d$ samples, and evaluate 
how anomalous finite-size effects could be exploited to detect this phenomenon experimentally in two-dimensional artificial systems, or in antiferromagnetic 
all-in--all-out pyrochlores like Nd$_2$Hf$_2$O$_7$ or Nd$_2$Zr$_2$O$_7$, for the first time.
\end{abstract}


\maketitle

\setlength{\textfloatsep}{10pt} 
\setlength{\intextsep}{10pt}

Order by disorder (ObD) is the mechanism whereby a system with a non-trivially degenerate ground state develops long-range 
order by the effect of classical or quantum fluctuations~\cite{Villain}. From a theoretical point of view, the ObD mechanism is a relatively common occurrence in geometrically frustrated spin models~\cite{Chalker}, such as the  fully frustrated domino model --where it was discussed for the first time~\cite{Villain}-- or the  Ising antiferromagnet on the three-dimensional FCC lattice~\cite{Lebowitz}. Many other theoretical realisations exist. However, definitive experimental 
evidence for this mechanism has remained elusive. Strong evidence for \emph{quantum} ObD in the antiferromagnetic XY insulating rare-earth pyrochlore oxide Er$_2$Ti$_2$O$_7$ has been reported~\cite{Champion03,Zhitomirsky12,Savary12,Ross14}, but a conclusive proof of
\emph{thermal} ObD remains unseen in the laboratory so far.  The difficulty lies in establishing whether order is selected through the ObD mechanism (a huge disproportion in the density of low-energy excitations associated with particular ground states) or it is due to energetic contributions not taken into account that actually lift the 
ground state degeneracy.

In this work we study ObD in Ising spin systems where the staggered order is inhibited by a magnetic 
field. We analyse theoretically and numerically the three-dimensional ($3d$) pyrochlore 
system and its two-dimensional ($2d$) projection (the checkerboard lattice). We demonstrate the existence of singular finite size effects (FSE) and we
show how they can be exploited to detect ObD. Our results suggest that thermal ObD could be finally observed experimentally in natural staggered structures based on the 
pyrochlores~\cite{Reimers91,Sadeghi15,Anand15} as well as in artificially designed two-dimensional magnetic~\cite{Marrows16} or colloidal systems~\cite{Reichhardt}.

More precisely, we first study an Ising pyrochlore with $\langle 111 \rangle$ anisotropy and antiferromagnetic (AF) nearest-neighbour
 interactions. In the absence of magnetic field (${\mathbf B}$), the ground state is the all-spins-in--all-spins-out
N\'eel state~\cite{Melko}. A strong field along the crystalline direction $[110]$ can break this order, turning it into a 
disordered state with three-spins-in/one-spin-out and three-out/one-in elementary units.  This type of disordered system of \emph{magnetic charges} (see below) had been studied before in the context of spin ice~\cite{Borzi13,Brooks14,Xie15}, but
in the presence of rather artificial constraints. In contrast, as we will show, the present case is obtained in a simple way, with the additional reward of exhibiting an ObD transition at moderate fields.  We give numerical evidence for this phenomenon and we prove it analytically with the cluster variational method (CVM)~\cite{Foini}, and a low-temperature analysis~\cite{Villain} of the $2d$ approximate projection on the checkerboard lattice
that allow us to exhibit singular FSE~\cite{Lukic}. We  explicitly show the relevance of the low-energy excitations on the ordering mechanism by evaluating the density of states of the $3d$ system. Finally,
we discuss the possibility to discriminate true ObD experimentally in three different scenarios. 

\begin{figure}[t]
\includegraphics[width=\linewidth]{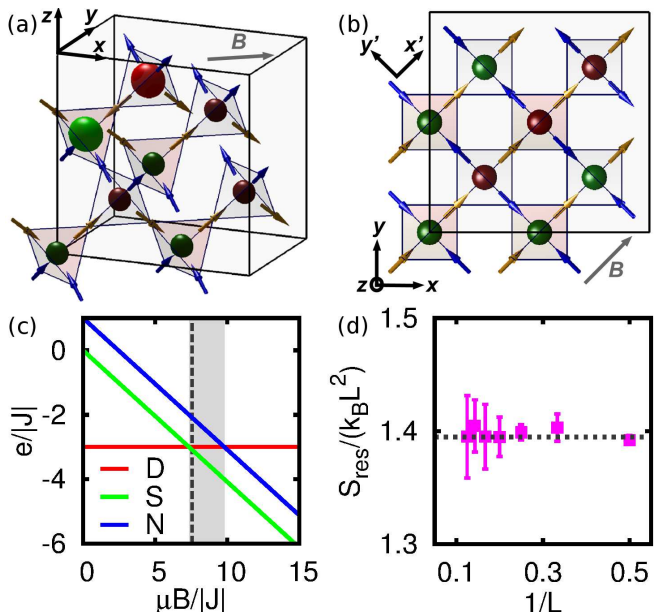}
\caption{\label{fig:uno}
(a) Conventional unit cell of the pyrochlore lattice ($L=1$) with $4$ tetahedra pointing up (coloured)
and $4$ pointing down (uncoloured).
The arrows represent the spin direction at
each site.
The colour of the spheres marks the sign of the
charges and the size is proportional to their modulus. (b)
Planar projected configuration: a tetrahedron becomes a square with crossings, in a checkerboard lattice. 
The field ${\mathbf B} \parallel [110]$
couples to two spins on each tetrahedron ($\alpha$-chains, yellow arrows), 
and is orthogonal to the other two ($\beta$-chains, blue
arrows). The configuration shown is a zero-temperature ground state for large fields.
(c) The energy of a singly charged (S),
doubly charged (D) and neutral tetrahedron (N) as functions of the magnetic energy normalised 
by the exchange energy scale; only the energies corresponding to the S and N configurations favoured by the field are shown.
(d) Linear size dependence of the residual entropy $S_{\rm res}$.}
\end{figure}

The pyrochlore lattice consists of corner-sharing tetrahedra, see Fig.~\ref{fig:uno}(a). 
The centres of tetrahedra pointing up (coloured) and down (uncoloured) make two interpenetrating FCC
lattices (a diamond lattice).
Classical Ising magnetic moments ${\boldsymbol \mu}_i =\mu {\mathbf S}_i  =\mu S_i \hat {\mathbf s}_i $ 
sit on the vertices of the tetrahedra. The quantisation directions $\hat {\mathbf s}_i$ are along the
$\langle 111 \rangle$ diagonals and, conventionally, $S_i = \pm 1$ indicates a magnetic moment 
pointing outwards or inwards of an up tetrahedron. The Hamiltonian is
\begin{equation}
{\mathcal H} = -J_{\textit{nn}}   \sum_{\langle ij\rangle} {\mathbf S}_i \cdot {\mathbf S}_j - \mu \sum_i {\mathbf B} \cdot {\mathbf S}_i
\; ,
\end{equation}
where $J_{\textit{nn}}$ is the exchange constant, 
and the first sum runs over nearest neighbours.
The ferromagnetic (FM) version of this Hamiltonian corresponds to the nearest-neighbour spin-ice model. Differently from other works, here we concentrate on the antiferromagnetic case. For ${\mathbf B} \parallel [110]$, one can understand the spin system as consisting of two types of chains:
while blue arrows ($\beta$-spins) belong to the  ``$\beta$-chains'' (running perpendicular to $\mathbf{B}$) with $\hat{\mathbf s}_i \cdot {\mathbf B} = 0$, yellow ones ($\alpha$-spins) 
sit on the ``$\alpha$-chains'' (parallel to $\mathbf{B}$), such that $ {\hat {\mathbf s}}_i \cdot {\mathbf B} = 
\alpha_i \sqrt{2/3} B$ and $\alpha_i = \pm 1$~\cite{Hiroi03}. Figure~\ref{fig:uno}(b) displays the conventional planar projection of the $3d$ lattice.
Using these definitions, one can rewrite the Hamiltonian in terms of scalar quantities:
\begin{equation}
{\mathcal H} = 
J \sum_{\langle ij\rangle} S_i S_j - \frac{ \sqrt{2}\mu B}{\sqrt{3}} \sum_{i\in \alpha} \alpha_i S_i
\; ,
\label{eq:dos}
\end{equation}
where $J < 0$ contains a geometrical factor (including a sign change)~\cite{Bramwell01} and the second sum runs over the $\alpha$-chains
only.  

Within the magnetic charge picture~\cite{Castelnovo}, the centres of up or down
tetrahedra are considered neutral if two of their spins
point in and two out, have a positive or negative single
charge if three spins point in and one out or vice
versa (pictured as small spheres in Fig.~\ref{fig:uno}(a,b)), or a
positive or negative double charge if all spins point in or all point out
(pictured as big spheres in Fig.~\ref{fig:uno}(a)). In this language, the ground state for
${\mathbf B} = 0$ consists of an array of double charges of alternating sign with the zincblende structure, 
which spontaneously breaks the symmetry between the two FCC sublattices~\cite{Guruciaga14}. Unlike the single charges and the neutral state,
a double charge has no magnetic moment, making it unstable under a sufficiently 
strong magnetic field applied along any direction. ${\mathbf B} \parallel [110]$ is 
special in that it does so without favouring any FCC sublattice~\footnote{${\mathbf B} \parallel  [100]$ also has this property, but it does not lead to
a single-charge ground state.}, opening the door to a single-charge disordered ground state. In order to measure the amount of charge order for a given spin configuration, we define the single and double staggered charge densities, $\rho_S^s$  and $\rho_S^d$ respectively. They represent the modulus of the magnetic charge density due to single or double monopoles in up tetrahedra, normalised so that full order corresponds to a value of 1. It is also useful to define  $\rho_S=\rho_S^s+2\rho_S^d$ representing the \emph{total} staggered charge per sublattice site.

As implied by Eq.~\eqref{eq:dos}, ${\mathbf B}$ lowers the
energy of single charges and neutral tetrahedra with
positive projection of magnetic moment along it, leaving that
of double charges  unchanged (see Fig.~\ref{fig:uno}(c)). A field $B$ 
such that $\mu B/|J| > 7.348$ stabilises a ground state which, while remaining globally neutral, has a single charge on
each tetrahedron. This field orders the
$\alpha$-chains ferromagnetically (Fig.~\ref{fig:uno}(b)), isolating the $\beta$-chains in the
same way that a $[111]$ field decouples the Kagom\'e planes
in the spin ice case~\cite{Sakakibara03}. Spins on $\beta$-chains are impervious 
to this field but not to the exchange interaction.
Each $\beta$-chain will thus independently and spontaneously 
order antiferromagnetically (see Fig.~\ref{fig:uno}(b)), implying a spontaneous one-dimensional staggered charge
order along each $\beta$-chain. The
additional freedom associated with the symmetry breaking within each separate
$\beta$-chain means that no $3d$
staggered charge order can arise
at $T = 0$, though the residual entropy (proportional
to the number of chains, not spins) is sub-extensive.
We tested this fact
using Monte Carlo (MC) simulations in a system of $L^3$ cubic cells and periodic boundary conditions  with an applied field
$\mu B/|J| = 12.1$, well in the disordered regime.
By integrating the specific heat over a wide temperature range
$(0.1 < k_B T/|J| < 70)$, and through direct calculation from
the density of states computed with the Wang-Landau (WL) algorithm~\cite{Wang}, we obtained the residual entropy 
$S_{\rm res} \propto L^2$ for different system
sizes (Fig.~\ref{fig:uno}(d)). Details on the simulations are provided in the Supplemental Material~\cite{supmat}.

Here we are interested in the situation in which
the ground state consists of this disordered single-charge state while the lowest energy excitations correspond to double charges (see the shaded area in Fig.~\ref{fig:uno}(c); the dashed line marks the field
$\mu B/|J| = 7.562$ used in the remainder of this work). The appearance of excitations implies an obvious entropy increase. On the other hand, their structure (all spins in, or all out) imposes nearest-neighbour correlations \cite{Guruciaga14} that will favour charge order between adjoining chains, and thus the phenomenon of ObD that we will study.

\begin{figure}[t]
\includegraphics[width=\linewidth]{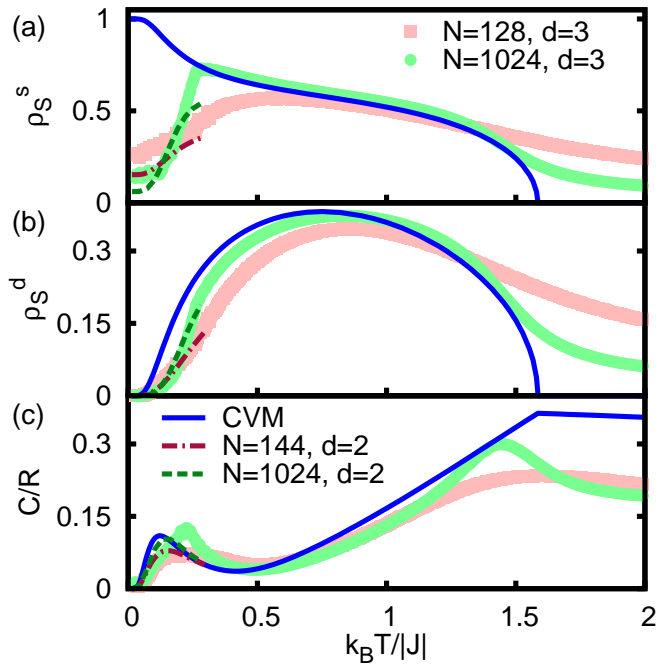}
\caption{\label{fig:dos}
The order parameters (a) $\rho_S^s$ and (b) $\rho_S^d$, and (c) the specific heat as functions of temperature. The symbols
show MC data for the $3d$ system and the dashed lines (with the same colour code)       
the results of the low-$T$ expansion for the $2d$ projection, both with similar
number of spins $N$. 
The solid (blue) lines are the CVM outcome ($N \to \infty$).
}
\end{figure}

The cluster variational method (CVM) places the tetrahedra on 
a tree with the same coordination number than the $3d$
lattice, i.e. four~\cite{Levis,Foini,Gonnella15} (Husimi tree).
Once this is done, recurrence relations for the order parameters
are derived and solved analytically in the infinite size limit. We tested the formation of charge order in the zincblende structure by measuring $\rho_S^d$ and $\rho_S^s$, plotted as continuous lines in Fig.~\ref{fig:dos}. Within the CVM both order parameters are strictly zero at $T=0$. At an infinitesimal temperature $\rho_S^s$ jumps to one and
$\rho_S^d$ increases continuously from zero. 
The two observables vanish at $T_c$ as in a second-order Ising-type phase transition with mean-field exponents.
The specific heat at low $T$ has a standard Schottky anomaly due to two-level system excitations. 
The second peak indicates the transition to the disordered phase and it is just a
cusp since $\alpha=0$ in mean-field. The MC simulations of the $3d$ model (symbols in Fig.~\ref{fig:dos}) clearly support this interpretation of the specific heat, with a low temperature Schottky anomaly at low $T$, and a broad peak with evident FSE at $T_c$ (Fig.~\ref{fig:dos}(c)). Its evolution with the number of spins $N$, as well as that of $\rho_S^s(T_c)$ and its fluctuations,
are consistent with a second-order transition within the $3d$ Ising universality class (see Supplemental Material~\cite{supmat}). 
In order to understand better the behaviour of the order parameter, we discuss the important low $T$ finite-size effects below.

A careful low-$T$ analysis is most easily implemented in the projection to the checkerboard lattice. 
As in the $3d$ case, the lowest energy levels are excited by flipping an $\alpha$-spin against $\bf{B}$. Furthermore, this $\alpha$-spin should link two $\beta$-chains with \emph{staggered} AF $1d$ order. Labeling the spins $s_{ij}$  in Fig.~\ref{fig:uno}(b) according to the coordinate system $(0; \, x, \, y)$, such 
an excitation would be to turn the spin $s_{22}$. This creates 
two defects with an energy cost equal to $2 \epsilon=2 \mu B-12 J$ ($\epsilon / J=0.164$ 
for the parameters of the numerical simulations). For simplicity, we used open 
boundary conditions in this representation. 

With the exact enumeration of these excitations we calculated 
$\rho_S^s$ and $\rho_S^d$ for finite $N$ at low $T$. We expect the 
FSE to depend on the number of spins $N$ 
on the lattice, independently of its dimension $d$. 
We see in Fig.~\ref{fig:dos} that the behaviour observed for $d=2$ qualitatively follows that of the numerical simulations of the $3d$ system. 
The naive extrapolation of $\rho_S^s$ in the low $T$ limit for $N\to\infty$ suggests that 
$\rho_S^s$ vanishes in the thermodynamic limit, in seeming
contradiction to the CVM results (top of Fig.~\ref{fig:dos}). To explain this, 
we provide below a careful study of the interplay between the limits $N\to\infty$ and $\beta\epsilon\to \infty$ 
in the checkerboard model.

Consider two neighbouring $\beta$-chains with $M$ spins in the direction $y'$ of the
rotated coordinate system $(0; x', y')$ in Fig.~\ref{fig:uno}(b)~\footnote{Note that the number of spins on two neighbouring columns in the
direction $y'$ may differ by $2$. However, for $M$ large enough, this
gives subleading boundary corrections which we will neglect in the
following.}.
In the low-$T$ limit the $\beta$-spins have perfect antiferromagnetic
order along the$y'$ direction.
There are thus two possibilities for the relative orientation of the spins
on neighbouring
$\beta$-chains: either they are parallel (FM ordered, e.g.,
the second and third $\beta$-chains in Fig.~\ref{fig:uno}(b)) or anti-parallel (AF
ordered, e.g., in the first and second $\beta$-chain in Fig.~\ref{fig:uno}(b)). In
the latter case and for $M$ large enough  there are $O(M/2)$  possible excitations of energy $2
\epsilon$,
obtained by reversing an  $\alpha$-spin between the two $\beta$-chains, and the partition function is
$Z_{AF} \simeq (1+e^{-2 \beta \epsilon})^{M/2}$. In the former case there
are no possible
low-energy excitations (neglecting the presence of neutral
tetrahedra) and $Z_{FM} \simeq 1$.
This can be interpreted in terms of an effective AF coupling between the $\beta$-spins only. Let us compare the partition functions for the two possible orientations,
\begin{equation}
\frac{Z_{\rm AF}}{Z_{\rm FM}} = e^{-\beta (H^{\rm AF}_{\rm eff} - H^{\rm FM}_{\rm eff})} 
\; , \quad
H_{\rm eff} =  J_{\rm eff} 
\sum_{j'=1}^{M}  s^{\beta}_{1j'} s^{\beta}_{2j'} 
\; , 
\end{equation}
where $s^{\beta}_{i'j'}$ label the $\beta$-spins and $i',j'$ sweep the rotated lattice  in Fig.~\ref{fig:uno}(b). The interaction on each $\beta$-chain remains the original $J$.
Then $H^{\rm AF}_{\rm eff} = - J_{\rm eff} M$, $H^{\rm FM}_{\rm eff} =  J_{\rm eff}M$
and  
we conclude that $J_{\rm eff}$ is given by 
\begin{equation}
J_{\rm eff} = (4\beta)^{-1} \ e^{-2\beta \epsilon}
\; . 
\end{equation}
Thus, after integrating out the $\alpha$-spins, we have an effective low-$T$ model
on a tilted $2d$ square lattice, with AF \emph{anisotropic} interactions
equal to $J$ in the $y'$ direction and $J_{\rm eff}$ in the $x'$ direction. 

We now go one step further and we reduce the low-$T$ effective model to a $1d$ one. As mentioned, the $\beta$-chains are perfectly ordered for $k_B T/|J| \ll 1$. We
define a macro $\beta$-spin $\sigma= +1$ ($-1$) according to the direction of the 
first spin on the chain being up (down). They sit on the sites of a $1d$ lattice and 
interact with an effective AF interaction $M J_{\rm eff}$.
The $T\to 0^+$ limit is then very tricky, since $J_{\rm eff}\to 0$ in a singular way. 
At any finite $M$, 
$M J_{\rm eff} \to 0$ for $T\to 0^+$: the effective $1d$ system decouples and disorders.
However, if we take $M\to\infty$ first, the effective $1d$ system orders AF 
and, as it was already AF ordered along the $\beta$-chains, it is fully AF-ordered with
$\rho_S^s \xrightarrow[T \to 0^+]{} 1$.
In conclusion, the low-$T$ expansion predicts a first-order ObD transition with a 
finite jump of $\rho_S^s$ from zero to $1$, just as obtained with the CVM. 
One has
\begin{equation}
0 = \lim_{N \to \infty} \lim_{T \to 0^+} \rho_S^s \neq \lim_{T \to 0^+} \lim_{N \to \infty} \rho_S^s = 1 \, .
\end{equation}
MC results are not in contradiction with 
this: huge values of $N$ are needed to see a large $\rho_S^s$ as $T$ lowers (Fig.~\ref{fig:dos}(a)).

\begin{figure}[t]
\includegraphics[width=\linewidth]{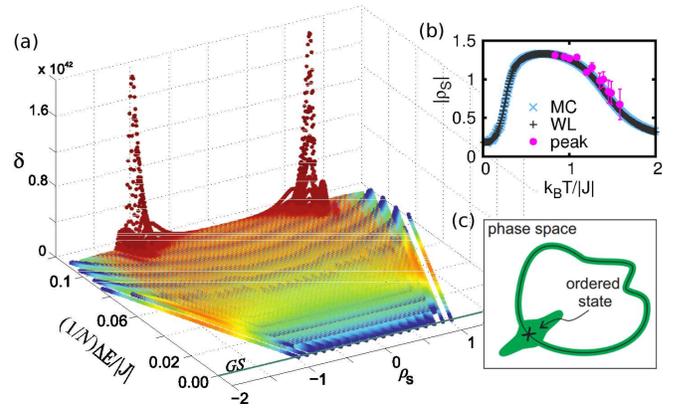}
\caption{\label{fig:tres}
(a) Density of states $\delta$ as a function
of excitation energy, $\Delta E$, and total staggered
charge density, $\rho_S$, for a system of size $L = 3$
and for $\mu B/|J| = 7.562$. Each coloured circle
corresponds to a single energy of the system. The two symmetric
peaks are centered at
$|\rho_S^{\rm peak}| \approx 1.2$.  Inset (b) compares $|\rho_S|$ calculated
from the WL method and the MC simulation with the values directly obtained from the maxima of $\delta$. Inset (c) --copied from Ref.~\cite{moessner2001magnets}-- represents the ground state manifold (black curve) in parameter space, and the selection of an ordered state (cross) by thermal excitations (green region).
}
\end{figure}

ObD is sometimes illustrated by a cartoon \cite{moessner2001magnets,Chalker} in which the ground state manifold is represented by a curve in phase space (see  
Fig.~\ref{fig:tres}(c$)$). The increment on the accessible phase space when raising $T$ slightly from $0$ is drawn as a surface (green) next to this curve. ObD occurs when the 
excitations linked to certain ordered ground states (cross) dominate the thermal average over all accessible states. We can turn this description into a quantitative argument by explicitly calculating the density of states $\delta(\Delta E,\rho_S)$
of the $3d$ system with the WL algorithm. We move along the configuration space using two parameters: the energy 
excess with respect to the ground state, $\Delta E$, and the total staggered charge density, $\rho_S$. Figure~\ref{fig:tres}
shows the low energy part of $\delta$ 
for $L = 3$ and the same magnetic field used in Fig.~\ref{fig:dos}. The colours emphasise
the value of $\delta$, normalised using that there is a single state with $\rho_S = 2$ (a double monopole crystal). Each
point in the graph represents the real binning in energy
and order parameter. 
We can see a very flat surface near the degenerate ground state energy 
($\Delta E = 0$) and two very noticeable symmetric peaks; {\it e.g.}, at $\Delta E/(N|J|) = 0.106$ the maxima are at 
$\rho_S^{\rm peak}\approx\pm 1.2$. Consistently, the MC canonical ensemble average of the order parameter and the one computed with
\begin{equation}
|\rho_S|^{\rm WL} = \frac{1}{{\cal Z}} \sum_{\Delta E_k}^{\Delta E_k^{
\rm max}} \sum_{{\rho_S}_l} |{\rho_S}_l| \ \delta(\Delta E_k,{\rho_S}_l) \ e^{-\beta \Delta E_k}
\label{eq:WL}
\end{equation}
with $\Delta E_k^{\rm max}/|J| = 0.385$ coincide over a rather  wide range of $T$,
see Fig.~\ref{fig:tres}(b). The remaining (pink) data-points are obtained as follows. We first read 
the density $\rho^{\rm peak}_S(\Delta E)$ that maximises $\delta$ for each $\Delta E$. Such an evaluation is
quite precise on the interval $\Delta E/(N|J|) \in [0.05,0.2]$ and it coincides, within numerical accuracy, with $|\rho_S|(\langle\Delta E\rangle)$ measured with the MC or WL methods (not shown). We next transform the 
$\Delta E$ dependence into a $T$ dependence replacing $\Delta E$ by $\langle \Delta E\rangle(T)$ from 
the MC data. We therefore obtain $|\rho_S^{\rm peak}|(T)$ in Fig.~\ref{fig:tres}(b). 
The good coincidence between the data points obtained in this way and through Eq.~\eqref{eq:WL} marks the importance of low-energy thermal fluctuations around the ordered states. Furthermore, the procedure complements the low-$T$ expansion in the sense that it estimates $|\rho_S|$ beyond its maximum as a function of $T$.

We will now discuss how an experimental 
contrast to our results may be possible with 
magnetic~\cite{Reimers91,Sadeghi15,Anand15,Xu15,Marrows16} or colloidal~\cite{Reichhardt} samples.
In order to observe classical ObD experimentally in these systems, the tendency towards a charge ordered ground state favoured by a misaligned ${\mathbf B}$ or long range 
dipolar interactions should be made as small as possible. 
A first option are the AF Ising pyrochlores~\cite{Reimers91,Sadeghi15,Anand15}.  They have the advantage that the stabilisation of the all-in--all-out state is achieved through 
the exchange interaction; this allows to keep the dipolar interaction small. Field misalignment can also be minimised by using a vector magnet \cite{Bruin2013}. The observation 
of charge order at temperatures much higher than that characterising both the residual field and dipolar energy scales would be a strong indication of ObD. 
Detecting the AF order disappear as a function of increasing field (see Fig.~\ref{fig:uno}(c)) 
would also be a conclusive smoking gun. Neutron scattering could be a sensitive probe if only short or medium 
range AF order is set due to competing ordering trends. 
A second route towards ObD is to take advantage of the recent advances in the design of frustrated magnets to create a planar system 
similar to the one in  Fig.~\ref{fig:uno}(b), thus minimising 
problems of field misalignment.
The idea is to use artificial square-lattice spin-ice samples in their AF phase, with the
vertex energy hierarchy $\epsilon_{\rm c} < \epsilon_{\rm e} < \epsilon_{\rm a,b}$ where ${\rm c}$ indicates AF, 
${\rm e}$ three-in/one-out or three-out/one-in and ${\rm a,b}$ FM vertices~\cite{Marrows16,Marrows} (see the Supplemental Material 
for a detailed explanation~\cite{supmat}). 
The third route concerns the use of colloidal systems in $2d$ arrays of optical traps, where staggered charge order could be attainable~\cite{Reichhardt,Tierno}. 
The precise control over the interactions and accessibility of thermal excitations may offer another fertile ground to recreate this experiment.
As we have seen, FSE have a very strong influence in $\rho_S^s$; this fact may be exploited to recognise incipient ObD using samples 
with small size.  

In summary, we have made an in-depth numerical and theoretical study of ObD in two models closely linked to ice systems. 
Our results may establish a route to the much sought-after experimental realisation of classical order-by-disorder.

\begin{acknowledgments}
We thank P. Holdsworth, L. Jaubert, C. Marrows and R. Moessner for
very useful discussions. This work was supported in part by MINCyT-ECOS A14E01,
PICS 506691 (CNRS-CONICET), NSF under Grant No. PHY11-25915,
ANPCyT through PICT 2013 N$^{\circ}$2004 and PICT 2014 N$^{\circ}$2618, and
Consejo Nacional de Investigaciones Cient\'{\i}ficas y T\'ecnicas (CONICET). 
MVF acknowledges partial financial support from Universidad Nacional de La Pampa,
Argentina. LFC is a member of the Institut Universitaire de France.
\end{acknowledgments}


%




\widetext
\clearpage

\begin{center}
\textbf{\large Supplemental Material: Field-tuned order by disorder in Ising frustrated magnets with antiferromagnetic interactions}
\end{center}

\section{Simulation methods and further analysis}

We provide here some details on the simulations used in the main text to study the equilibrium properties of nearest-neighbour antiferromagnetic Ising pyrochlores. We simulated $L\times L \times L$ conventional cubic cells of the pyrochlore lattice with periodic boundary conditions in the three directions with Metropolis and Wang-Landau~\cite{Wang} methods. In both cases we used a single-spin flip algorithm. In the Metropolis case, after reaching equilibrium we averaged the data over $10-150$ independent runs and $5 \times 10^4 - 1.5 \times 10^6$ time steps depending on temperature and lattice size. 
We have used a modified version of the Wang Landau Algorithm proposed
by R.~Bellardinelli and V.~Pereyra~\cite{belardinelli2007} to compute
the density of states $\delta (\Delta E, \rho_S) $ as a function of
the order parameter $\rho_S$ and the excitation energy, restricting
this last one to a range $\Delta E / (N |J|) = 0.385$ from the ground
state. The time-dependent modification factor changed from $e^1$ to
$e^{10^{-7}}$ evolving as a function of the inverse of the Monte Carlo
time.  The final result is a relative density of state of the system.
We used the condition $\delta(\rho_s =2) =1$ (where the addition over
all energies is implied) to normalise it.

\begin{figure}[htb]
\includegraphics[width=0.5\linewidth]{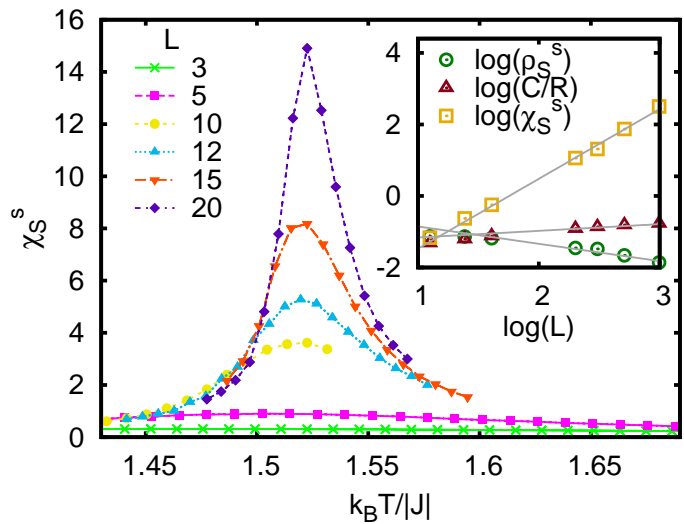}
\caption{\label{smfig:uno}
The single charge susceptibility $\chi_S^s$ displays finite
size effects consistent with a second order phase transition. Inset: $\rho_S^s\left(T_c \left(L\right)\right)$, 
$C\left(T_c \left(L\right)\right)/R$ and 
$\chi_S^s\left(T_c \left(L\right)\right)$ evolve as power laws with the size of the system, and present good correspondence with the behavior expected 
for the three dimensional Ising model universality class  (gray lines).}
\end{figure}

The main text is concerned with the order-by-disorder (ObD) process at low temperature; we now pay some attention to the high temperature feature (the usual ``disorder-by-disorder'' transition). In this case, while order is reinforced by the increasing number of double charges at higher $T$, the proliferation of neutral excitations has the opposite effect of decorrelating magnetic charges~\cite{Guruciaga14}, thus disordering the system. To study this process, we performed a finite size analysis on systems with $L$ between $3$ and $20$ (a maximum of $128000$ spins). We show in Fig.~\ref{smfig:uno} the single charge susceptibility $\chi_S^s$ (defined as the quadratic fluctuation of $\rho_S^s$ over temperature) in the temperature range of this transition. This quantity, as well as the specific heat $C$ and $\rho_S^s$ at the critical point, evolve with $L$ as in a
second order phase transition (not shown). The critical exponents we observe are consistent with the three-dimensional Ising universality class (Fig.\
\ref{smfig:uno}, inset).

\section{Details on the antiferromagnetic ground state for the artificial spin ice}

In this Supplemental Material we explain how ObD can be realized in artificial spin-ice (ASI) samples  on planar square lattices. Differently from the three dimensional pyrochlores, deviations of the field orientation will not have 
drastic effects in these systems, making them suitable for the observation of thermal ObD.

\begin{figure}[tb]
\includegraphics[width=0.5\linewidth]{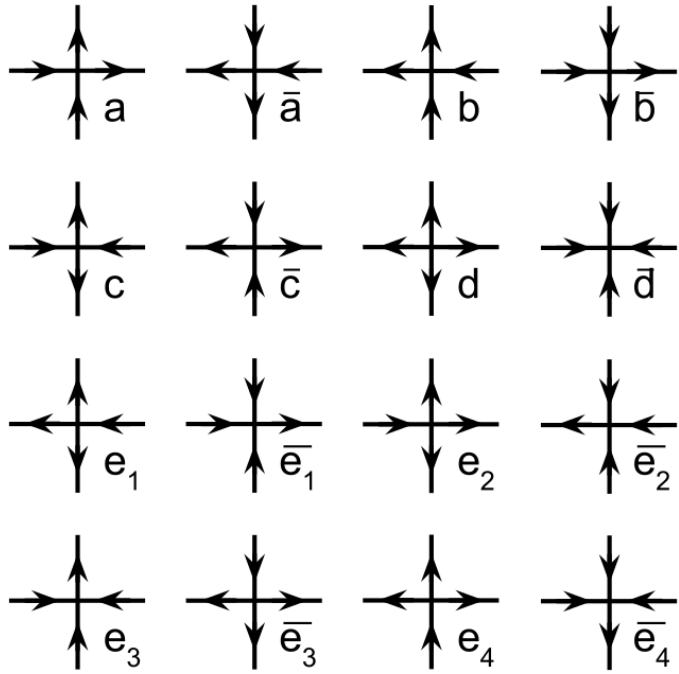}
\caption{The sixteen vertices.}
\label{smfig:dos}
\end{figure}

ASI materials are  made of arrays of elongated single-domain
ferromagnetic nano-islands frustrated by dipolar interactions.
The interaction parameters can be precisely engineered -- by tuning
the distance between islands, i.e. the lattice constant, or by
applying external fields. These
systems should set into different phases depending on the experimental conditions.
Despite the irrelevance of temperature fluctuations, 
methods to equilibrate ASI up to quite low temperatures have been devised. Two of them are the gradual
magneto-fluidization of an initially polarized state and the thermalization during the slow growth of
the samples~\cite{Marrows}.
Neglecting dipolar interactions between the magnetic islands, these systems can be described with 
the sixteen vertex model on a square lattice, with individual vertices shown and labeled as in Fig.~\ref{smfig:dos}~\cite{Levis,Foini}.

The idea is to engineer samples with the following (unusual but feasible~\cite{Marrows16}) vertex energy hierarchy:

\begin{equation}
\epsilon_{\rm c} < \epsilon_{\rm e} < \epsilon_{\rm a,b} < \epsilon_{\rm d} \; . 
\end{equation}

\begin{figure}[tbh]
\includegraphics[width=0.5\linewidth]{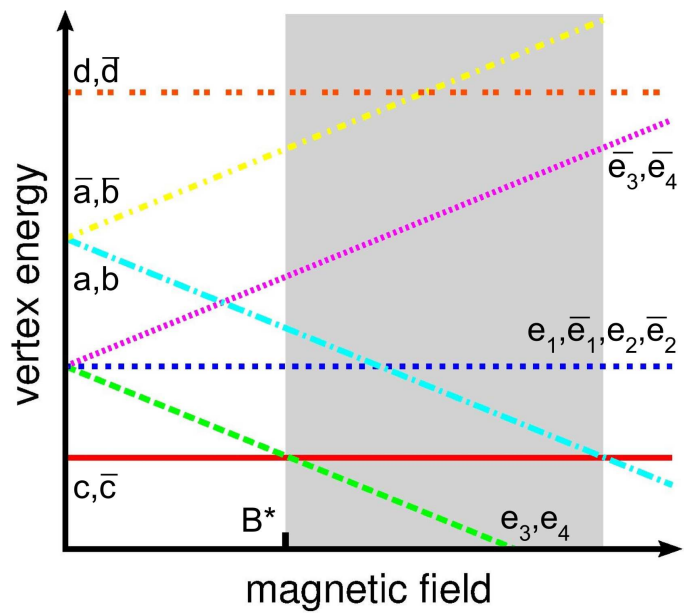}
\caption{The dependence of the vertex energies on the applied magnetic field in the 
vertical direction. As in Fig.~1 in the main text, the shaded rectangle marks the field range where order-by-disorder would be expected.}
\label{smfig:tres}
\end{figure}

\noindent The ground state of the system is the two-degenerate fully antiferromagnetic state, with alternating ${\rm c}, \ \overline {\rm c}$ vertices.
If one applies a magnetic field with magnitude $B$ in the $y$ direction, all vertices with vertical arrows pointing upwards 
decrease their energy (${\rm a}, \ {\rm b}, \ {\rm e}_3, \ {\rm e}_4$), complementary all vertices with vertical arrows pointing 
downwards increase their energy ($\overline {\rm a}, \ \overline {\rm b}, \ \overline {\rm e}_3, \ \overline {\rm e}_4$) while 
all other vertices are insensitive to this field. The vertex energy dependence on the field is 
sketched in Fig.~\ref{smfig:tres}. For $B>B^*$ the energies of the ${\rm e}_3, \ {\rm e}_4$ vertices cross the ones 
of the ${\rm c}, \ \overline {\rm c}$ vertices and the ground state becomes a disordered state made of ${\rm e}_3, \ {\rm e}_4$
vertices. Indeed, all vertical arrows must point in the direction of the field. Therefore, each horizontal line
is a sequence of alternating ${\rm e}_3$ and ${\rm e}_4$ vertices but there is no constraint on the 
way two consecutive horizontal lines order relative to each other. The number of ground states is proportional to $2^L$, with $L$ the number of lines. 

\begin{figure}[tb]
\includegraphics[width=0.5\linewidth]{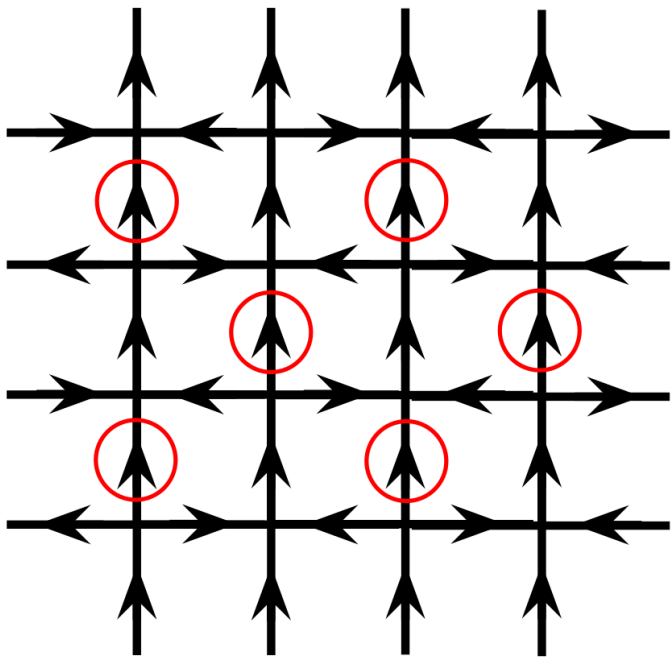}
\caption{One of the two ground states with the most numerous low energy excitations. 
We indicate with circles the spins that could be flipped with the lowest energy cost, the only possible at very low temperatures.}
\label{smfig:cuatro}
\end{figure}

If we choose a field with strength slightly larger than $B^*$, by the same mechanism explained in the 
main text, two ground states with staggered order between the horizontal lines (see Fig.~\ref{smfig:cuatro} where one of them is shown) have a huge number of low energy
excitations that correspond to flipping all the encircled vertical arrows that transform
the ${\rm e}_3$ and ${\rm e}_4$ vertices into ${\rm c}$ and $\overline {\rm c}$ vertices. This feature produces 
an effective antiferromagnetic interaction between the horizontal spins that 
orders the system as soon as these excitation are activated by temperature. An 
effective one dimensional model can be derived for this problem following the 
same steps presented in the main text. Strong finite size effects are also present in this 
case and we propose to use them as a signature for ObD by analysing experimental
data at low, but not so low, temperatures for moderate system sizes.
Note that this scenario is not destroyed by a slight misalignment of the field. However, 
the dipolar interactions could lift the degeneracy between the ground states. Care must then be 
taken to make the dipolar interactions as weak as possible.

\end{document}